# A facile and low environmental impact approach to prepare thermally conductive nanocomposites based on polylactide and graphite nanoplatelets

*Alberto Fina[1]\*, Samuele Colonna[1], Lorenza Maddalena[1], Mauro Tortello[1], Orietta Monticelli[2]\**

[1]Dipartimento di Scienza Applicata e Tecnologia, Politecnico di Torino-sede di Alessandria, viale Teresa Michel, 5, 15121 Alessandria, Italy

[2]Dipartimento di Chimica e Chimica Industriale, Università di Genova, Via Dodecaneso, 31, 16146 Genova, Italy

alberto.fina@polito.it

orietta.monticelli@unige.it

**ABSTRACT**. In this work, the preparation of nanocomposites based on poly(L-lactide) PLLA and graphite nanoplatelets (GNP) was assessed, by applying, for the first time, the reactive extrusion (REX) polymerization approach, which is considered a low environmental impact method to prepare polymer systems and which allows an easy scalability. In particular, *ad hoc* synthesized molecules, constituted by a pyrene end group and a poly (D-lactide) (PDLA) chain (Pyr-D), capable of interacting with the surface of GNP layers as well as forming stereoblocks during the ring opening polymerization (ROP) of L-lactide, were used. The nanocomposites were synthesized by adding to L-lactide the GNP/initiator system, prepared by dispersing the graphite in the acetone/Pyr-D solution, which was dried after the sonication process. DSC and X-ray diffraction measurements evidenced the stereocomplexation of the systems synthesized by using the pyrene-based initiators, whose extent turned out to depend on the PDLA chain length. All the prepared nanocomposites - including those synthesized starting from a classical initiator, that is 1-dodecanol - retained similar electrical conductivity, whereas the thermal conductivity was found to increase in the stereocomplexed samples. Preferential localization of stereocomplexed PLA close to the interface with GNP was demonstrated by Scanning Probe Microscopy (SPM) techniques, supporting for an important role of local crystallinity in the thermal conductivity of the nanocomposites.

**Introduction**

Improving the properties and expanding the application fields of polymers from renewable origin, such as polylactide (PLA), which is the object of the present work, represent a significant challenge to make these materials a real alternative to fossil based polymers.[1] With this aim, different fillers/nanofillers, such as silica,[2] layered silicate,[3] POSS[4] and graphite,[5-11] were previously combined with PLA. The addition of carbon fillers/nanofillers turned out to improve the crystallization rate of the polymer,[5-8] these acting as nucleating agent,[9] to enhance mechanical and gas barrier properties[8,10] and to confer electrical conductivity[11,12] to the polymer matrix. A key aspect for the development of the above materials is related to the preparation method, which should be easy, low cost and characterized by a low environmental impact. Typically, the preparation of PLA/graphite systems was carried out by using the solution-mixing approach,[10,12] which implies the preliminary dispersion of the layered carbon filler in a solvent able to promote exfoliation and to disperse the graphene layers as well as to solubilize the polymer. Although this approach was found to guarantee a fine nanoflakes dispersion,[10-12] it is based on the use of high environmental impact solvents, such as chloroform or dichloromethane. On the other hand, melt processing is well known to be an industrially viable method for the preparation of polymer nanocomposites but yet very challenging for the dispersion of carbon nanoflakes in PLA.[13]

Moreover, in order to promote specific interactions of the filler with the polymer, graphene oxide (GO), which contains hydroxyl and epoxy groups on the basal planes and carboxyl groups on the edges, was used.[14] The exploitation of GO requires the oxidation of graphite in harsh conditions, typically performed by using strong acids and oxidating agents,[14] thus achieving the extensive modification of the carbon atoms hybridization.

It is worth underlining that, to restore the electrical and thermal conductivity of pristine graphene, subsequent reduction of GO has to be performed.[14] However, complete reduction and healing of the defective structure is extremely difficult and requires the use of strong chemical reducing agents and/or extremely high temperature. On these bases, the development of novel approaches, combining the use of nanoflakes obtained via environmental friendly routes with sustainable and scalable processing into polymer nanocomposites is of great interest. In particular, the reactive extrusion (REX) is currently one of the most promising way to synthesize and modify PLA, as well as to prepare composite/nanocomposites, which technique allows to perform simultaneously the polymerization from lactide and the dispersion of nanoparticles.[15-20] Indeed, since the pioneering work,[15] which demonstrated the feasibility of the method in the ring opening polymerization of lactide, REX was applied for the synthesis and chemical modification of PLA-based materials in reactions such as coupling from PLA precursors,[16] free-radical grafting of PLA chains[17] and transesterification.[18] Concerning the preparation of composite/nanocomposite systems, Bourbigot et al.[19] applied reactive extrusion to develop PLA/carbon nanotubes composites aiming at enhancing flame retardancy. It is worth underling that the polymerization time used in the above process (ca. 50 min) was much higher than that normally acceptable for REX processes, which is in the order of minutes. Another nanofiller used to produce composites based on PLA, by applying the REX process, is layered silicate. In this case, Nishida et al.[20] carried out the preparation of PLA/clay nanocomposites following a two-step procedure, namely a mixing process, which lasted for almost 90 minutes and allowed the polymer intercalation among the clay layers, followed by an extrusion process, which enhanced the filler exfoliation. While these pioneering works demonstrated the feasibility of REX in the preparation of composites, significant development in the chemical design of the system is needed to make the process sustainable and industrially viable, especially in terms of processing time and efficiency of nanoparticles dispersion.

In order to apply the reactive extrusion to the preparation of PLA/GNP nanocomposites and optimizing the material final properties, *ad hoc* synthesized oligomers were used as initiators of the ring opening polymerization of L-lactide. Indeed, these molecules, easily synthesized in bulk, starting from the commercial 1-pyrenemethanol (Pyr-OH), are constituted by a short chain of PDLA and a pyrene end group (Pyr-D). The exploitation of such initiators is expected to promote from one side specific interactions of pyrene terminal groups with the surface of graphite layers[21] and from the other allow to obtain stereoblock systems, whose structuration might occur close to the graphite surface (**Figure 1**). It is worth underling that the stereocomplexation, which is generated by the strong interactions between L-lactyl and D-lactyl unit sequences, can improve the properties of PLA-based materials.[22,23] Although PLA stereocomplex systems have been intensively investigated,[22,23] the influence of the stereocomplexation on the polymer thermally conductive properties was not previously reported to the best of the authors' knowledge. In fact, high interest in currently focused on thermally conductive polymer composites for exploitation in several engineering applications, including low temperature heat recovery and heat storage, heat exchangers for corrosive environment and flexible heat spreaders.[24-26]

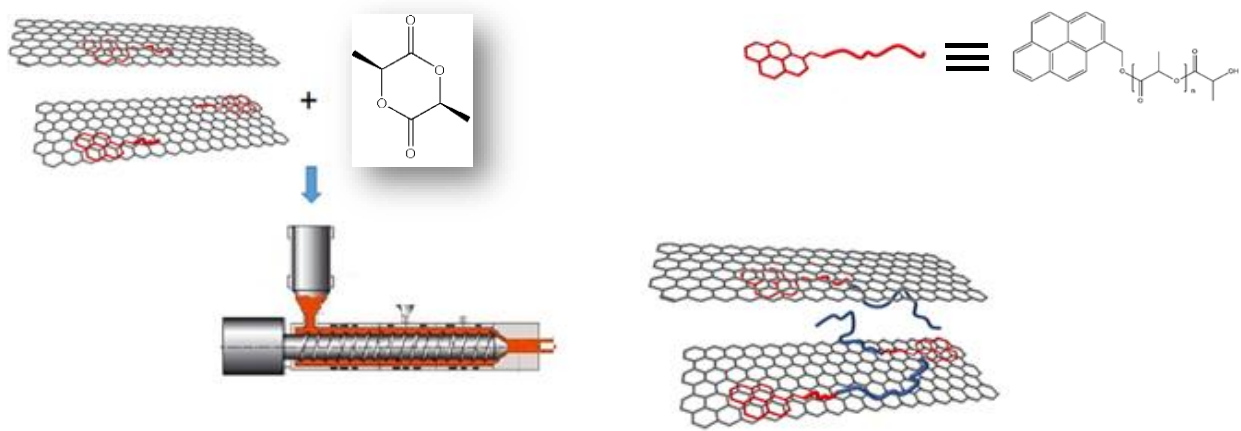

**Figure 1**. Scheme of the nanocomposites preparation procedure.

**Materials and methods**

**Materials.** D-lactide (D-la) and L-lactide (L-la) (purity >98%) were kindly supplied by Purac Biochem (The Netherlands). Before polymerization, the monomers were purified by three successive recrystallizations from 100% (w/v) solution in anhydrous toluene and dried under vacuum at room temperature. Tin(II) 2-ethylhexanoate ($Sn(Oct)_2$) (95%; from Sigma-Aldrich), 1-pyremethanol (Pyr-OH) and 1-dodecanol, from Sigma-Aldrich, were used without further treatments. All the solvents (*i.e.*, anhydrous toluene (≥99.7%), methanol and acetone) were purchased from Sigma-Aldrich and used as received. Graphite nanoplatelets (GNP, G2Nan grade, **Figure 1S**), with lateral size ~ 10÷50 μm, thickness ~ 10 nm, Surface area ≈ 40 $m^2$ $g^{-1}$, Raman $I_D/I_G$ ≈ 0.06, Oxygen content ≈ 1.7 at.% (XPS, O1s signal), were kindly supplied by Nanesa (I) and used as received.

**Pyr-D synthesis.** Pyrene-based oligomers, named Pry-D, were synthesized by ring-opening polymerization (ROP) of D-lactide, using Pyr-OH as initiator and $Sn(Oct)_2$ as a catalyst, in bulk at 120 °C.[27] The details about the synthesis are reported in the Supporting Information.

**Preparation of PLLA-based systems by reactive extrusion.** Different initiators, both commercial and *ad hoc* synthesized, were used in the ROP of L-la by reactive extrusion. Indeed, 1-dodecanol and two Pyr-D oligomers, with different average molar masses, were exploited in the above polymerization and the resultant samples were coded as follows: PLLA_DOD_R (PLLA prepared in the extruder, by using 1-dodecanol as initiator), PLLA_Pyr_2500_R (PLLA prepared in the extruder, by using a pyrene-based oligomer with molar mass of 2500 g/mol as initiator) and PLLA_Pyr_8000_R (PLLA prepared in the extruder, by using a pyrene-based oligomer with molar mass of 8000 g/mol as initiator). The sample prepared by applying a common batch reactor was defined with the letter "B", such as PLLA_DOD_B (PLLA prepared in the batch reactor, by using 1-dodecanol as initiator). Moreover, for the nanocomposites, the letter "G" was added to the above described codes. The initiator concentration in the reaction mixture was adjusted in order to obtain the same molar mass for all the synthesized samples. For the preparation of the nanocomposites, an appropriate amount of GNP, to obtain 5 wt.-% in the final product, was first dispersed in anhydrous acetone containing the initiator, by sonication in a sonic bath (Ney Ultrasonic) at 40 Hz for 1 hour. The above mixture, dried in vacuum at room temperature to completely remove the solvent, was added to a previous purified L-la. The same procedure was followed in the preparation of the neat systems. The polymers were prepared by melt mixing the lactide/initiator or lactide/initiator/GNP mixture into a co-rotating twin screw micro-extruder (DSM Xplore 15, Netherlands) for 5 minutes at 100 rpm and 230°C. Then, $Sn(Oct)_2$ ([L-la]/[$Sn(Oct)_2$] = $10^3$) was added and the process was carried out (constant screw speed and temperature) until the decrease of the melt viscosity, which occurred after ca. 5 minutes (see details in the Results and Discussion part).

**Characterization.** FTIR spectra were recorded on a Bruker IFS66 spectrometer in the spectral range 400-4000 $cm^{-1}$.

$^1$H-NMR spectra were collected with a Varian NMR Mercury Plus instrument, at a frequency of 300 MHz, in $CDCl_3$ solutions containing tetramethylsilane as internal standard. Pyr-D samples were analyzed by applying this technique.

The GPC analysis was performed on THF solutions using a 590 Waters chromatograph equipped with refractive index and ultraviolet detectors and using a column set consisting of Waters HSPgel HR3 and HR4 with a flow rate of 0.5 mL $min^{-1}$. The column set was calibrated against standard PS samples.

A Zeiss Supra 40 VP field emission scanning electron microscope (FE-SEM) equipped with a backscattered electron detector was used to examine the composite morphologies. The samples were submerged in liquid nitrogen for 30 min and fractured cryogenically. Nanocomposites were sputter-coated with a thin carbon layer using a Polaron E5100 sputter coater. GNP were deposited on $SiO_2$/Si and observed without any further preparation.

Differential scanning calorimetry (DSC) measurements were performed with a Mettler-Toledo TC10A calorimeter calibrated with high purity indium and operating under flow of nitrogen. The sample weight was about 5 mg and a scanning rate of 10 °C/min was employed in all the runs. The samples were heated from 25 °C to 230 °C, at which temperature the melt was allowed to relax for 1 minute, then cooled down to -10 °C, and finally heated up again to 230 °C (second heating scan).

Static wide-angle X-ray diffraction was carried out in reflection mode a Philips PW 1830 powder diffractometer (Ni-filtered Cu Ka radiation, k = 0.1542 nm).

Electrical conductivity (volumetric) was measured on disk-shape specimens (thickness and diameter of 1 and 25 mm, respectively) prepared by compression molding with a homemade apparatus described in previous works.[25,28] The conductivity value was calculated with the following formula:

$$\sigma = \frac{1}{\rho} = \frac{1}{\frac{V}{I} \cdot \frac{S}{l}} \left[\frac{S}{m}\right] \qquad (1)$$

where $S$ and $l$ are the specimen surface and thickness, respectively; $V$ is the voltage and $I$ the electric current, both read by the apparatus.

Isotropic thermal conductivity tests were carried out on a TPS 2500S by Hot Disk AB (Sweden) with a Kapton sensor (radius 3.189 mm) on disk-shaped specimens (prepared by compression molding of dried nanocomposites) with thickness and diameter of about 4 and 15 mm, respectively. Before each measurement, specimens were further stored in a constant climate chamber (Binder KBF 240, Germany) at 23.0 ± 0.1 °C and 50.0 ± 0.1 %R.H. for at least 48 h before tests. The test temperature (23.00 ± 0.01 °C) was controlled by a silicon oil bath (Haake A40, Thermo Scientific Inc., USA) equipped with a temperature controller (Haake AC200, Thermo Scientific Inc., USA).

SPM measurements were carried out on an Innova microscope from Bruker. The VITA module was used for the collection of nano-thermal analysis (Nano-TA) curves. FMM measurements were carried out using FESPA-V2 probes. Detailed description of SPM measurements is reported in the Supporting Information.

**Results and discussion**

The preparation of nanocomposites based on poly(L-lactide) (PLLA) and graphite nanoplatelets (GNP) was first concentrated on the synthesis of oligomers, made of a poly(D-lactide) (PDLA) chain attached to a pyrene end group, to be further used as initiators of the L-lactide (L-la) polymerization accomplished by reactive extrusion (REX). Indeed, the ROP of D-lactide (D-la), employing 1-pyremethanol (Pyr-OH) as the initiator, was carried out, as described in the Experimental Section, by using a simple bulk method, thus avoiding the exploitation of solvents. The FTIR spectra of the crude reaction products (not shown) were recorded to verify that the conversion of D-la was close to completion. After purification, the polymerization products were characterized by means of $^1$H-NMR spectroscopy (see **Figure 2S**) and the signals identified according to the literature.[29] The mean degree of polymerization, based on $^1$H-NMR spectroscopy, was calculated by comparison of the peak integral of the methine protons in the polylactide chain with those at the chain end (at δ 5.16 and δ 4.35 ppm, respectively). The molecular masses of the oligomers (2500 g/mol and 8000 g/mol) calculated by $^1$H-NMR, turned out to be in satisfactory agreement with the theoretical values, thus demonstrating a fine control of the polymerization reaction (see Supporting Information).

In order to verify the feasibility of the polymerization reaction by applying the reactive extrusion process, the characteristics of the polymers synthesized by using the latter approach (PLLA_DOD_R) and the classical bulk polymerization (PLLA_DOD_B) were compared (**Table 1S**). The synthesis were carried out employing the same conditions and 1-dodecanol as initiator as well as Sn(Oct)$_2$ as catalyst, without adding other additives, in order to maintain the process as simple as possible. In particular, in the case of the REX method, the torque required for screws rotation was recorded as a function of time and the reaction was stopped when the former started to decrease, *i.e.* at the beginning of the polymer scission. As shown in **Figure 3S**, by applying the conditions reported in the Experimental Section, the decrement of the torque occurred approx. 5 minutes after the addition of the catalyst. It is worth underling that the same torque trend was found for all the prepared polymers and nanocomposites. Indeed, the application of the above residence times (<10 min) allows to figure out a feasible exploitation of the polymerization process.

Complete monomer conversion (**Table 1S**) was obtained for PLLA_DOD_R whereas the presence of the peak characteristic of the monomer at 936 cm$^{-1}$ in the FT-IR spectrum (**Figure 4S**) for PLLA_DOD_B evidence for the incomplete conversion. This observation is in agreement with previous literature reports,[13] and it is explained by the limited diffusion of the monomer in the static glass reactor, whereas a more efficient mixture is expected during melt compounding in the microextruder. Similar spectra were obtained for the other samples prepared by applying the reactive extrusion.

The molar masses of the synthesized polymers, (Table 1S), were found to be higher for the sample prepared by the bulk process than that synthesized by REX, $M_n$ being ca. $50 \cdot 10^3$ g·mol$^{-1}$ and $30 \cdot 10^3$ g·mol$^{-1}$ for PLLA_DOD_B and PLLA_DOD_R, respectively. This finding is likely related to the absorption of humidity by lactide during the reactive processing in the compounder. Nevertheless, the molecular weight distribution (MWD) turned out to be slightly narrower in the sample prepared in the extruder, namely 1.3 and 1.5 for PLLA_DOD_R and PLLA_DOD_B, respectively. It is worth underling that the value of MWD calculated for PLLA_DOD_R is even smaller than those generally reported in the literature for PLA prepared by reactive extrusion.[15] It is possible to infer that the reduced residence time as well as the exploitation of initiator molecules allow controlling the polymerization, which occurs with limited intermolecular transesterification reactions, although in our system stabilising agents were not used. Moreover, very similar molar masses were found for all the synthesized samples, owing to the initiator concentration which allowed to get the same degree of polymerization.

The direct linkage of PLLA molecules to the pyrene-based initiators, constituted by a PDLA chain, was proved by washing the samples with acetone, namely a solvent for the Pyr-D oligomers but not for PLLA. The amount extracted from the pyrene-based samples was about 1-2 wt.-% with respect to the starting mass, that is a quantity lower than the amount of initiator added to the reaction mixture and which may be also due to the removal of the catalyst.

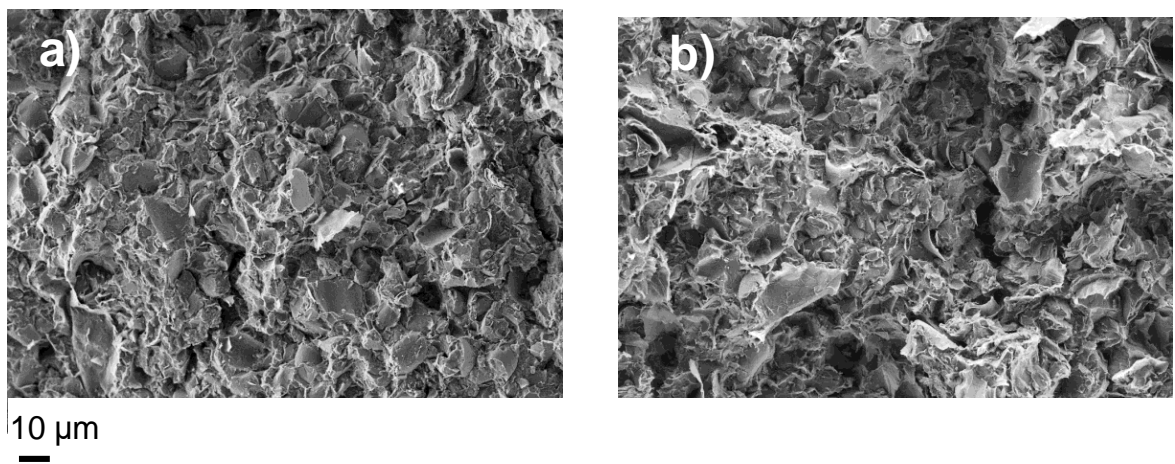

**Figure 2**. SEM micrographs of: a) PLLA_DOD_R_G and b) PLLA_Pyr_2500_R_G

The dispersion of GNP in the nanocomposites was evaluated by means of FE-SEM analysis. The comparison of the micrographs of the cryogenically fractured nanocomposites, PLLA_DOD_R_G (**Figure 2a**) and PLLA_Pyr_2500_R_G (**Figure 2b**), evidences in both the samples a homogenous distribution of GNP particles, whose flake dimensions are in the range of 10 µm or less. It is worth underlining that the other nanocomposite sample, prepared by using the Pyr_8000, had a very similar morphology. The above results demonstrated the applicability of the used approach for the preparation of nanocomposites characterized by a uniform distribution of the filler.

The thermal properties of the prepared composites were compared with those of the neat PLLA samples (**Figure 3**). **Table 1** summarizes all the relevant data extracted from the DSC measurements. DSC traces of the PLLA samples, shown in Figure 3A and related to the second heating scan, evidence, for all the analysed systems, a melting endotherm at ca. 170 °C, which corresponds to the melting of homocrystals. Furthermore, the samples prepared by using the pyrene-based initiators, together with the above melting peak, show another endotherm at higher temperature (around 220 °C), whose amount turned out to be affected by the PDLA chain length,

ΔHm$_S$ being 2 J·g$^{-1}$ and 16 J·g$^{-1}$ for PLLA_Pyr_2500_R and PLLA_Pyr_8000_R, respectively, clearly reflecting the longer PDLA block. Indeed, the above peak, which is ascribable to the melting of stereocomplex crystallites, demonstrated the structuring of the Pyr-D chain with those of PLLA. The addition of GNP to the reaction mixture was found to affect the thermal properties of the resultant materials. In all the samples, an increase of the crystallization temperature, more relevant in the case of PLLA_DOD_R_G and PLLA_Pyr_2500_R_G, occurs. This finding evidences that the filler acts as a nucleating agent for the crystallization of the polymer system. Moreover, it is worth underling, especially in the case of the sample PLLA_Pyr_2500_R_G, the specific nucleation effect of GNP on the stereocomplex crystal formation, which phenomenon has been already reported in the literature.[6] Nevertheless, comparing the DSC data, it is clear that all the nanocomposites are characterized by a similar overall crystallinity, which is around 60 %. Therefore, effects on conductivity by volume exclusion, previously proposed for stereocomplexed PLLA/PDLA/CNT nanocomposites[30] are not expected in this work. The detail data of the sample crystallinity is given in **Table 2S**. The crystalline structure was also studied by means of WAXD measurements. The patterns of the sample prepared by using 1-dodecanol, both neat and with GNP, shows the typical peaks of the α-form (at 2θ of 15°, 17° and 19°) of homocrystalline PLLA.[22] In the case of the systems based on pyrene oligomers, in addition to the above signals, peaks at 2θ of 12°, 22° and 24° appear, which are characteristic of the stereocomplex crystallites.[22] These findings, together with DSC results, indicate the partial stereocomplexation of the samples based on pyrene initiators.

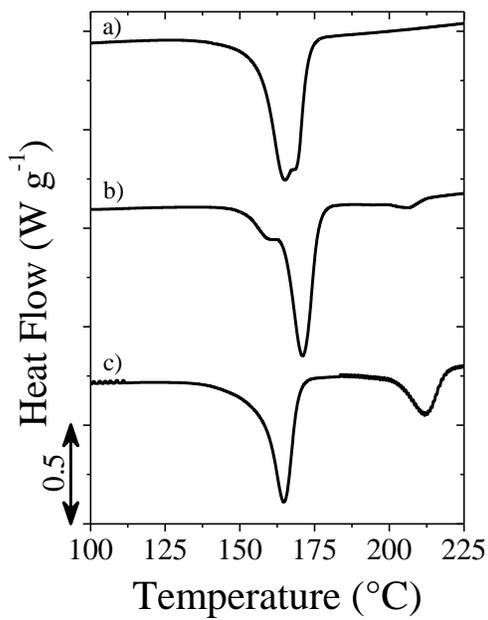
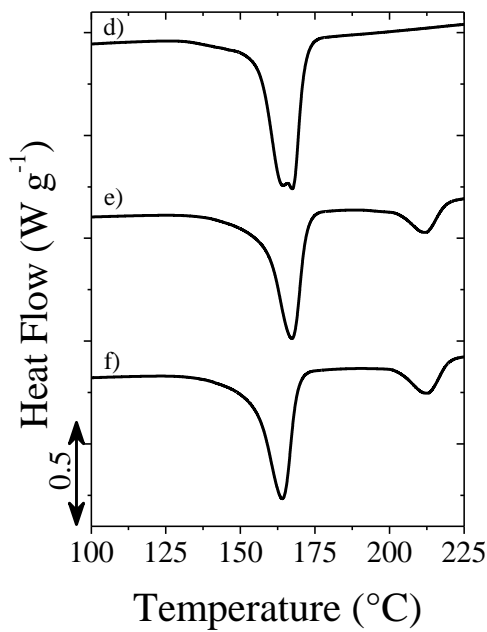

**A)**

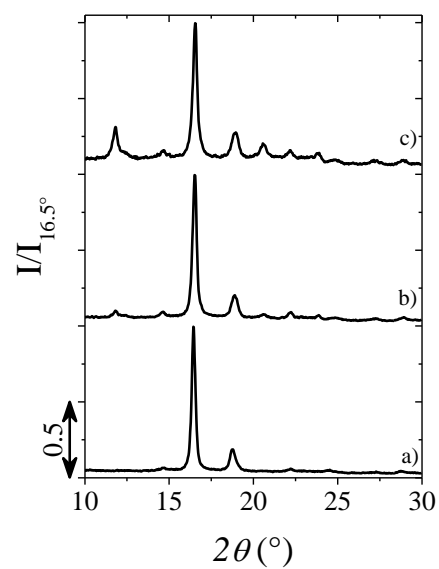
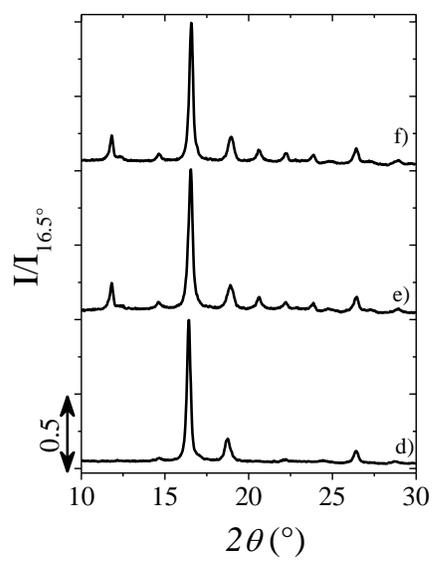

**B)**

**Figure 3**. A) DSC traces of: a) PLLA_DOD_R, b) PLLA_Pyr_2500_R, c) PLLA_Pyr_8000_R, d) PLLA_DOD_R_G, e) PLLA_Pyr_2500_R_G and f) PLLA_Pyr_8000_R_G; B) WAXD profiles of: a) PLLA_DOD_R, b) PLLA_Pyr_2500_R, c) PLLA_Pyr_8000_R, d) PLLA_DOD_R_G, e) PLLA_Pyr_2500_R_G, and f) PLLA_Pyr_8000_R_G.

**Table 1**. Characteristics of the prepared samples.

| Sample code | $T_c$ [°C] | $T_m$ [°C] | $\Delta H_m$ [J·g$^{-1}$] | $\Delta H_{m_s}$ [J·g$^{-1}$] | $\sigma$ [S·m$^{-1}$] | $\lambda$ [W·m$^{-1}$K$^{-1}$] |
|---|---|---|---|---|---|---|
| PLLA_DOD_R | 103 | 165 | 57 | - | $\cong 10^{-12}$ | 0.23 |
| PLLA_DOD_R_G | 113 | 167 | 58 | - | $1.8 \cdot 10^{-4}$ | 0.78 |
| PLLA_Pyr_2500_R | 97 | 171; 207 | 47 | 2 | $\cong 10^{-12}$ | 0.24 |
| PLLA_Pyr_2500_R_G | 123 | 167; 212 | 45 | 10 | $9 \cdot 10^{-5}$ | 0.85 |
| PLLA_Pyr_8000_R | 124 | 165; 212 | 40 | 16 | $\cong 10^{-12}$ | 0.24 |
| PLLA_Pyr_8000_R_G | 125 | 164; 213 | 47 | 12 | $4.6 \cdot 10^{-4}$ | 0.94 |

The subscripts m and c indicate the values measured during melting and crystallization, respectively. ΔH is the enthalpy, normalized to the PLLA content. The subscript $m_s$ indicates the value measured for the stereocomplexed fraction.

The effect of GNP on the electrical properties of the polymer system was studied by comparing the conductivity (σ) of the neat PLLA with those of the nanocomposites (Table 1). While PLLA is electrically insulating with an electrical conductivity in the range of $10^{-12}$ S·m$^{-1}$, the addition of GNP to PLLA leads to a dramatic increase in the electrical conductivity of the nanocomposites, achieving values in the range of ~ $10^{-4}$ S·m$^{-1}$, thus evidencing the formation of a percolating network.

In addition to the electrical conductivity, PLLA and its nanocomposites were also analysed in terms of thermal conductivity and the results are shown in the Table 1. The thermal conductivity ($\lambda$) of the neat PLLA is 0.23 W m$^{-1}$ K$^{-1}$, in agreement with values recently reported in the literature.[31,32] Furthermore, the presence of stereocomplexed PLA in PLLA_Pyr_2500_R and PLLA_Pyr_8000_R does not significantly change $\lambda$ of the polymer, which is coherent with the similar total crystallinity. Indeed, thermal conductivity is known to primarily depend on the total crystallinity of the polymer,[26,33] while different polymer crystalline forms are expected to have minor effect on the thermal conductivity, as long as the total crystallinity is below 70-80%.

As expected, the addition of GNP to the polymer matrix turns out to significantly influence the thermal conductivity, with values in the range 0.78÷0.94 W m$^{-1}$ K$^{-1}$. These values are significantly higher compared to results previously reported in the literature for both graphite[32] and graphite nanoplatelets[34] at the same wt.-% used in this work. Most interestingly, the thermal conductivity of the nanocomposites strongly depends on the type of initiator, with significantly enhanced thermal conductivities in the presence of stereocomplexed PLLA. Indeed, while PLLA_DOD_R_G exhibited a thermal conductivity $\lambda \approx 0.78$ W m$^{-1}$ K$^{-1}$, values of ~ 0.85 and ~ 0.94 W m$^{-1}$ K$^{-1}$ where measured for PLLA_Pyr_2500_R_G and PLLA_Pyr_8000_R_G, respectively, thus suggesting a role of the stereocomplex on the heat transfer between GNP and PLLA.

In order to explain the peculiar behaviour of our prepared systems, some aspects related to the thermal conductivity of composites have to be considered. In general, the thermal conductivity is determined by the structure and properties of both the polymer and the fillers, the dispersion and distribution of particles within the (nano)composites and the interactions between polymer chains and conductive particles, as well as between particles within a percolating network.[26] Pyrene-terminated PLA chains may indeed act as compatibilizers between the conductive GNP particles and the polymer matrix, which may enhance the thermal conductivity by different possible phenomena. While the presence of a compabilizer may affect the degree of dispersion of GNP, the thermal resistance associated to the GNP-polymer interfaces may also be reduced as a consequence of polymer-GNP non covalent bonding, resulting in a better heat transfer between GNP and the polymer matrix. Finally, given the presence of a dense percolating network of GNP witnessed by the electrical conductivity of all nanocomposites, contact thermal resistance between partially overlapping GNP flakes may also play an important role in the overall thermal conduction.[26] In particular, the presence of a highly ordered polymer layer between the GNP flakes may in principle result in a higher efficiency of heat transfer between nearby GNP flakes, taking into account that highly crystalline polymers are known to reach significantly higher thermal conductivity.[26] Indeed, pyrene-terminated compatibilizers able to self-organizestereocomplexed PLA domains were designed on purpose, aiming at the controlled organization of crystalline domains close to the surface of GNP. In fact, both Pyr_2500 and Pyr_8000 were found effective in increasing thermal conductivity in the presence of GNP, compared to the reference PLLA_DOD_R_G. Despite similar total crystallinity was found in PLLA_Pyr_2500_R_G and PLLA_Pyr_8000_R_G, a difference in thermal conductivity was observed, that might be related to difference in local organization of crystals onto the GNP. To investigate the formation of stereocomplex and to gain insight in its spatial organization within the composite, SPM analyses were performed on cryo-cut surfaces. Dispersion and distribution of GNP was evaluated on for PLLA_DOD_R_G and PLLA_Pyr_8000_R_G (**Figure 5S**), showing a fair distribution of flakes, with thickness in the range of ten nanometers and lateral size in the range of a few microns, in agreement with FESEM analyses reported above. Furthermore, NanoTA allows allowed monitoring the local thermal expansion as a function of temperature, thus providing a nanoscale characterization of the thermal properties. Nano-TA was indeed validated as a sensitive nanoscale method to verify formation of stereocomplex (see Supporting Information for details, **Figure 6S**). Nano-TA analyses were carried out on the nanocomposites surfaces in several points, either a) in close proximity (<100 nm) to nanoflakes oriented perpendicularly to the surface, representative of the properties of the PLA/GNP interfacial region, or b) at distance > 1µm from the nanoflakes, reflecting properties of the PLA matrix (**Figure 4**). In PLLA_DOD_R_G, for both points close and far from the interface with GNP flakes, M-shaped plots were observed, related to the occurrence of cold crystallization in the range 80-100°C, while softening and penetration of the probe occurs above 130°C, consistently with semicrystalline PLLA (see Supporting Information). No significant difference was observed in the average probe penetration temperatures close to the interface (141±9°C) and far from it (142±7°C), suggesting no modification of the crystalline organization of PLA at the interface with GNP. Deflection vs. temperature curves for the polymer matrix in PLLA_Pyr_8000_R_G show a penetration temperature (143±5°C) in the same range as for PLLA_DOD_R_G, but the plots display a different shape, with partially overlapping features related to limited cold crystallization and polymer softening, suggesting incomplete stereocomplexation of the polymer, in agreement with DSC and XRD analyses. On the other hand, deflection vs. temperature curves recorded in regions close to the interface with GNP show a higher penetration temperature (152±3°C) and no clear evidence of cold crystallization, both facts clearly supporting a higher degree of stereocomplexation of PLA in the interfacial region.

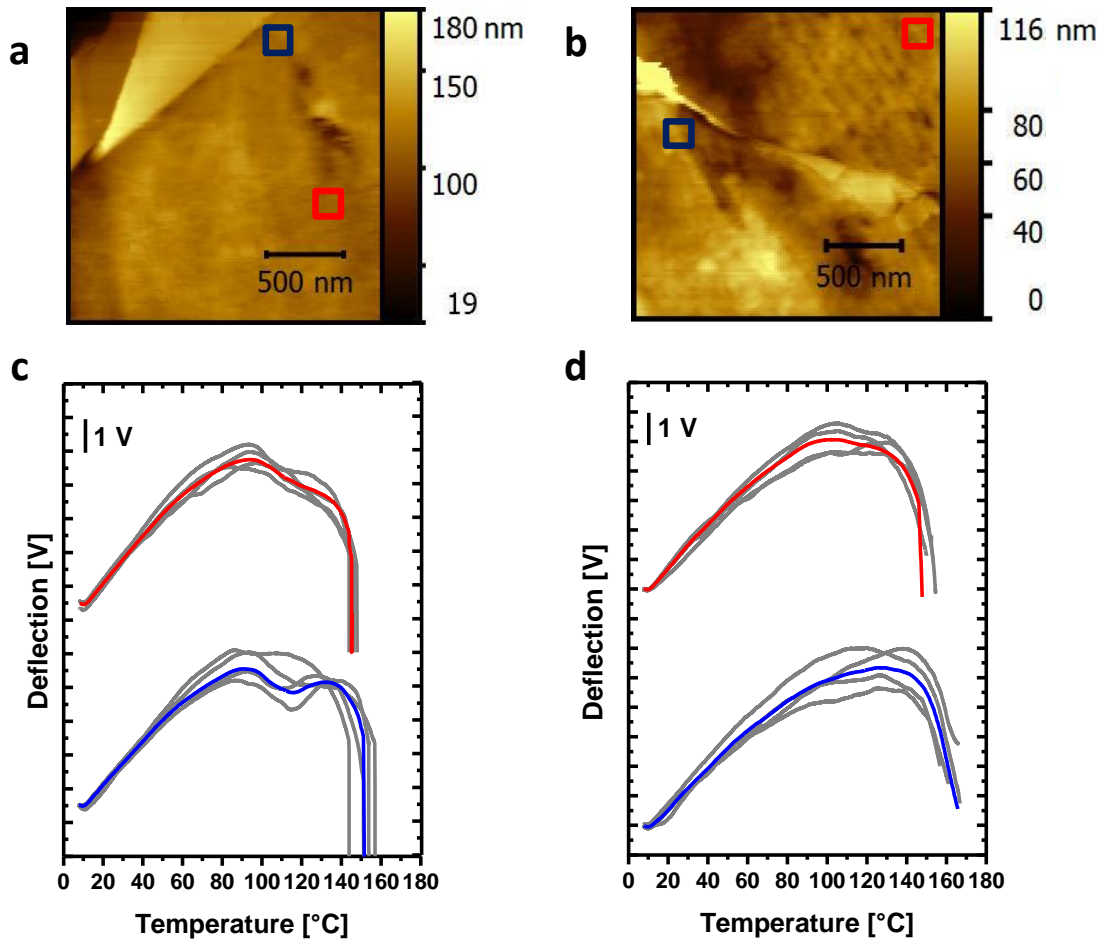

**Figure 4**. Representative topography maps for PLLA_DOD_R_G a) and PLLA_Pyr_8000_R_G b), showing the presence of GNP flakes approx. perpendicular to the cryo-cut surfaces. Red and blue squares indicate points of Nano-TA analyses far from the GNP and at the interface, respectively. Nano-TA deflection vs. temperature for PLLA_DOD_R_G c) and PLLA_Pyr_8000_R_G d) reports 4 tests performed in different points, for both areas close to the interface of far from it, as well as their average plots. Plots at the interface or far form the flakes are vertically-shifted for clarity.

To further investigate the difference in thermal properties of the interfacial region in PLLA_Pyr_8000_R_G, FMM measurements were performed, to investigate the viscoelastic properties of the polymer located close to the interface, compared to those of the bulk polymer matrix. In **Figure 5**, a relatively large GNP flake is clearly visible, since it is folded onto the surface as a consequence of the cryo-cut. In the interfacial area, both phase and amplitude signals show a difference in the contrast with respect to the polymeric matrix further away from the interface with the flake. In particular, the phase signal in the first 300 nm from the GNP flake appears to be significantly different compared to the bulk polymer. This result is consistent with nano-TA results and further support for a significant difference in crystalline organization of the polymer at the interface with GNP.

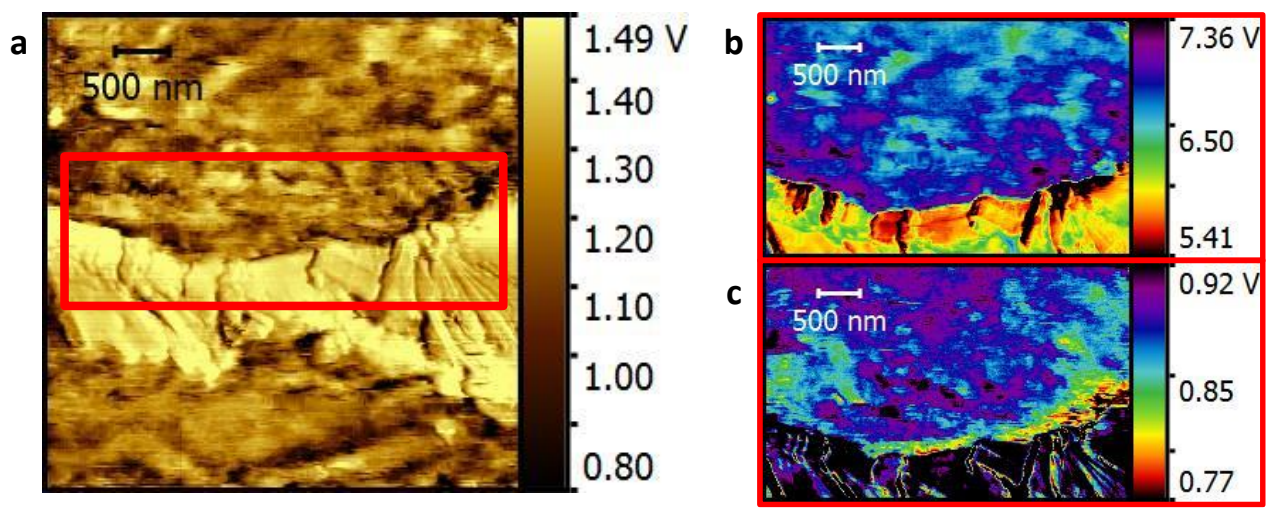

**Figure 5**. Lateral Signal Map of a graphene flake in PLLA_Pyr_8000_R_G a); FMM amplitude b) and FMM phase c) signal.

NanoTA and FMM results indicate a higher crystalline organization at the interface in the case of PLLA_Pyr_8000_R_G and confirm the role of Pyr-D in the compatibilization between GNP and PLA as well as in the local organization of the stereocomplex. The higher degree of stereocomplexation close to GNP in the presence of PyrD is indeed a proof of the effective grafting of pyrene-terminated oligomers to GNP. Furthermore, the high local crystallinity in the interfacial regions may explain the enhanced thermal conductivity performance of PLLA_Pyr_8000_R_G compared to PLLA_DOD_R_G, by the reduction of thermal resistances a the interface between GNP and the polymer as well as contact thermal resistance between overlapping nanoflakes.

**Conclusions**

This work demonstrated the viability of the reactive extrusion (REX), a low environmental impact as well as easily scalable approach, to obtain thermally and electrically conductive nanocomposites based on poly(L-lactide) PLLA and graphite nanoplatelets (GNP). The GNP dispersion, the tuning of the polymer molecular mass and its structuring, which occurs through the formation of stereoblocks, was affected by the exploitation, as polymerization initiators, of *ad hoc* synthesized oligomers made of a poly(D-lactide) (PDLA) chain attached to a pyrene end group. Stereocomplexation of the macromolecular chains was proven by SPM techniques to occur preferentially close to the surface of GNP, for the specific interactions of the pyrene group of the polymerization initiator with the nanofiller. The controlled organization of stereocomplexed domains turned out to significantly enhance the thermal conductivity of the nanocomposites, likely reducing the interfacial thermal resistances in the nanocomposite structure. The above material features, together with the easy and sustainable preparation procedure, open the bio-based nanocomposites to several low-temperature heat management applications, including heat storage and recovery, heat exchangers for corrosive environment and flexible heat spreaders.

**Supporting Information**. FE-SEM of graphite nanoplatelets, details of pyrene-based initiators synthesis and SPM methods, $^1$H-NMR of an initiator together with a table reporting the characteristics of the synthesized pyrene-based initiators, torque as a function of time during the reactive extrusion polymerization of the sample PLLA_DOD_R, table reporting the characteristics of the prepared polymers, FTIR spectra of the polymers prepared by applying REX and batch, AFM images for dispersed GNP flakes within the polymer, Nano-TA deflection vs temperature plots for PLA-based systems.

**Corresponding Author**

A. Fina and O. Monticelli are both corresponding authors.

alberto.fina@polito.it

orietta.monticelli@unige.it

**Acknowledgement.** This work was partially funded by the European Research Council (ERC) under the European Union's Horizon 2020 research and innovation programme grant agreement 639495 — INTHERM — ERC-2014-STG. The authors are grateful to PURAC BIOCHEM (The Netherlands) for supplying the lactide monomers and Nanesa (Italy) for supplying GNP. Dr. Maria Del Mar Bernal Ortega is gratefully acknowledged for helpful discussions on various aspects of the present work.

**For Table of Contents Use Only**

*Ad hoc* synthesized pyrene-based molecules allow to enhance the thermal conductivity of PLA/GNP nanocomposites prepared by the environmental friendly reactive extrusion

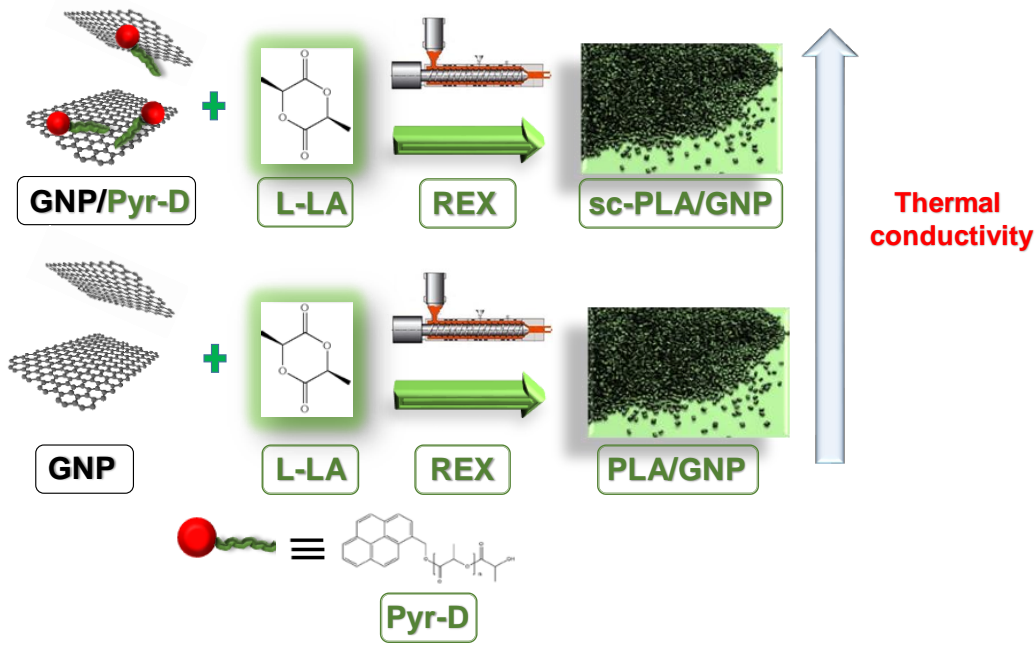

**Supporting information**

# A facile and low environmental impact approach to prepare thermally conductive nanocomposites based on polylactide and graphite nanoplatelets

*Alberto Fina, Samuele Colonna, Lorenza Maddalena, Mauro Tortello, Orietta Monticelli*

Page numbers: 11

Figure numbers: 6

Table numbers: 2

S1

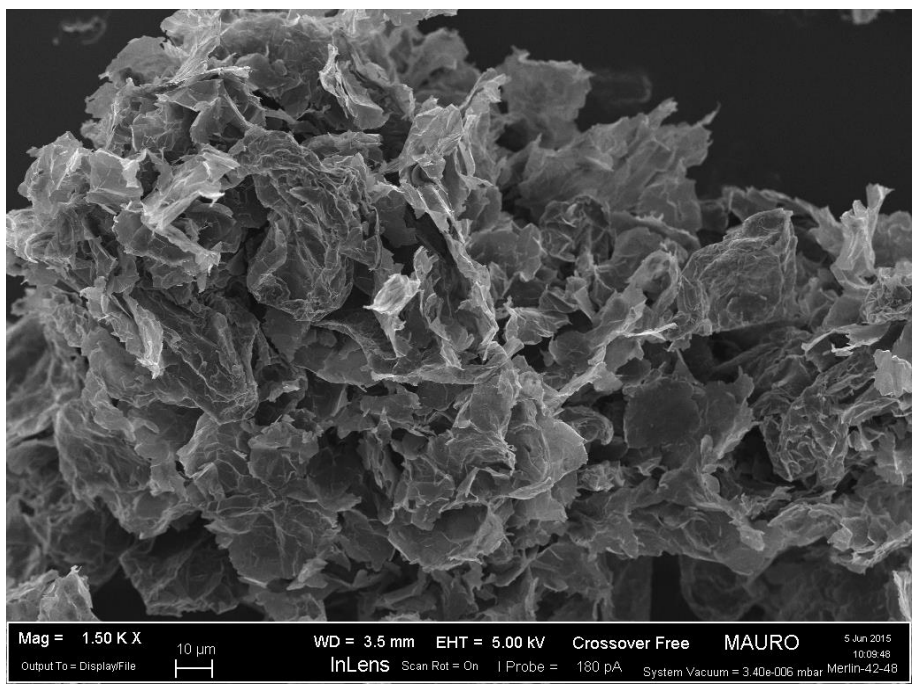

**Figure S1**. FE-SEM of graphite nanoplatelets G2Nan grade from Nanesa (I).

**Pyr-D synthesis**. In detail, about 10 g of D-la (previously recrystallized and dried) were charged under argon flow into the reactor (*i.e.*, a 50-mL two-neck round-bottom flask equipped with a magnetic stirrer) containing a predetermined amount of accurately weighed and in-situ dried Pyr-OH. The feeding ratio of the D-la monomer to the initiator was adjusted to obtain different molecular masses. After the introduction of D-la the flask was evacuated for 15 min and purged with argon, and the exhausting/refilling process repeated thrice in order to fully dry the reaction environment. The reaction vessel was then immersed into a thermostatically controlled oil bath set at 120 °C, under vigorous stirring: as soon as the mixture was completely molten and homogenized, about 0.15 mL of a freshly prepared solution of Sn(Oct)$_2$ in toluene ([D-la]/[Sn(Oct)$_2$] = 10$^3$) were added under argon through a micro-pipette, and the reaction allowed to proceed for 24 hours under inert atmosphere.

The crude product was then cooled down, dissolved in toluene and poured into an excess of cold methanol: after filtering and drying in vacuum at 40 °C to constant weight, the as-purified oligomer was obtained as a fine, yellowish powder.

**Scanning Probe Microscopy methods**. The VITA-HE-NANOTA-200 contact mode Si probes were adopted which feature a resonant frequency of about 55-80 kHz, a spring constant of 0.5-3 N/m and a tip radius < 30 nm. Moreover, in these probes there is a resistive heater integrated at the end of the cantilever, which can be repeatedly and reliably heated up to 350 °C. Thus, after having recorded standard contact-mode topography images, the probe can be placed in correspondence of the regions of interest (domains, features, phases etc.), for performing localized NanoTA. In a NanoTA measurement, the chosen location is heated by the probe and temperature-dependent properties, such as softening and melting, are investigated by recording the cantilever deflection vs temperature curves. The probes are calibrated by using three standards with known melting points, i.e. PCL (Tm=55°C), PE (Tm=116°C) and PET (Tm=235°C): when these samples are heated, the deflection increases due to the thermal expansion until it suddenly drops at the input power that corresponds to that specific melting temperature. The softening temperatures were calculated by fitting linearly the closest interval before and after the sudden variation of the deflection slope.

FMM measurements were carried out using FESPA-V2 probes that feature a nominal resonant frequency of 75 kHz and a spring constant of 2.8 N/m. In FMM, the probe oscillates in contact with the sample and the mechanical properties of the material under study can be investigated and compared by detecting both force amplitude and phase. The oscillation frequency was optimized in order to maximize the contrast in the amplitude and phase signals. The topography and lateral signals were also simultaneously recorded.

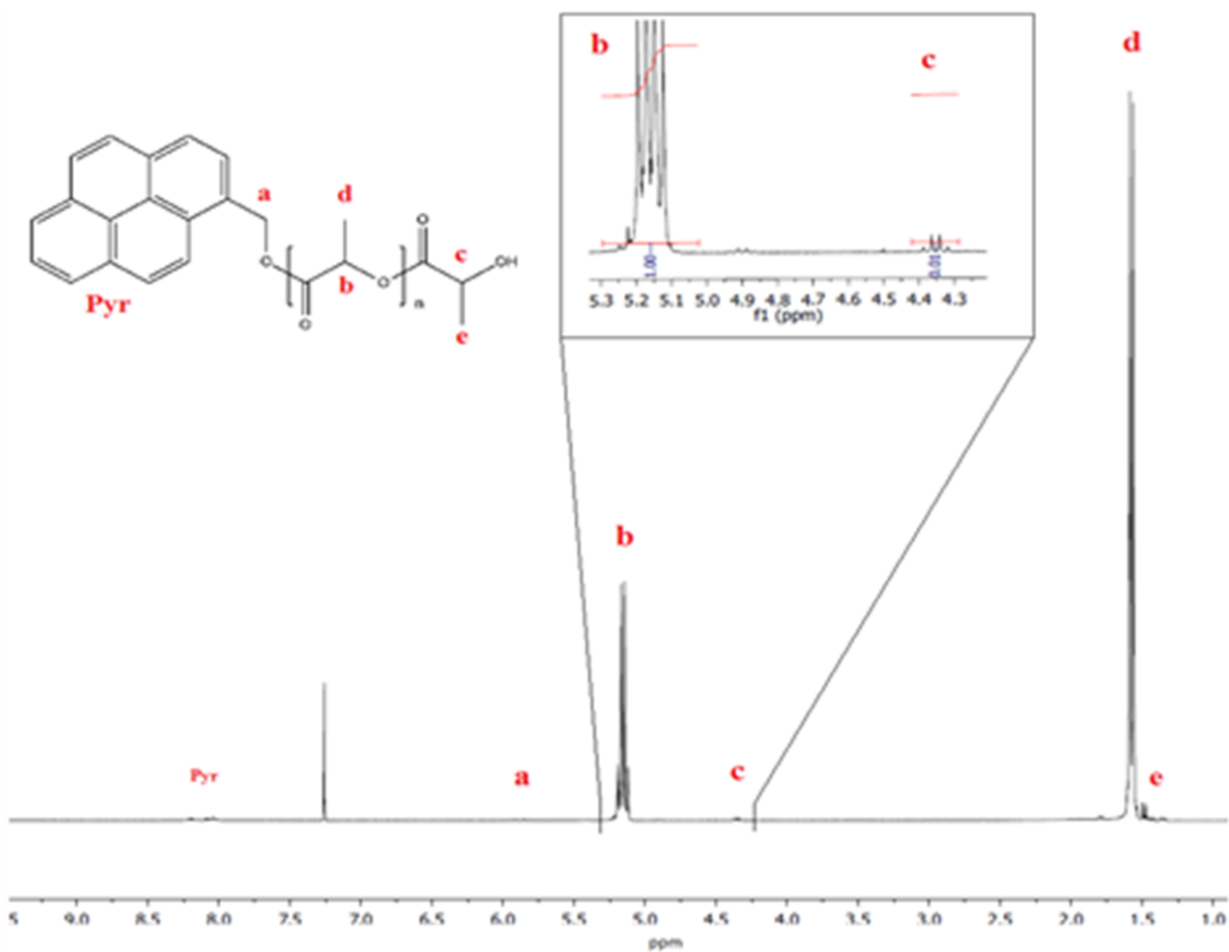

**Figure S2**. $^1$H-NMR of Pyr-D_8000.

| Sample | DP$_{NMR}$$^a$ | M$_{nNMR}$$^b$ [g/mol] |
|---|---|---|
| Pyr-D_2500 | 17 | 2500 |
| Pyr-D_8000 | 54 | 7800 |

$^a$Average degree of polymerization of D-la evaluated from $^1$H-NMR spectroscopy [by comparing the peak integral of the methine protons in the chain (at δ 5.16 ppm) with that of the methine protons at the chain end (at δ 4.35 ppm), according to the equation: DP$_{NMR}$ = 1/2 · (I$_{5.16}$ + I$_{4.35}$)/I$_{4.35}$]. $^b$Numeric average molecular weight of Pyr-D calculated as: Mn$_{NMR}$ = DP$_{NMR}$ · 144.13 + M$_{Pyr-OH}$, where 144.13 is the molecular mass of D-la.

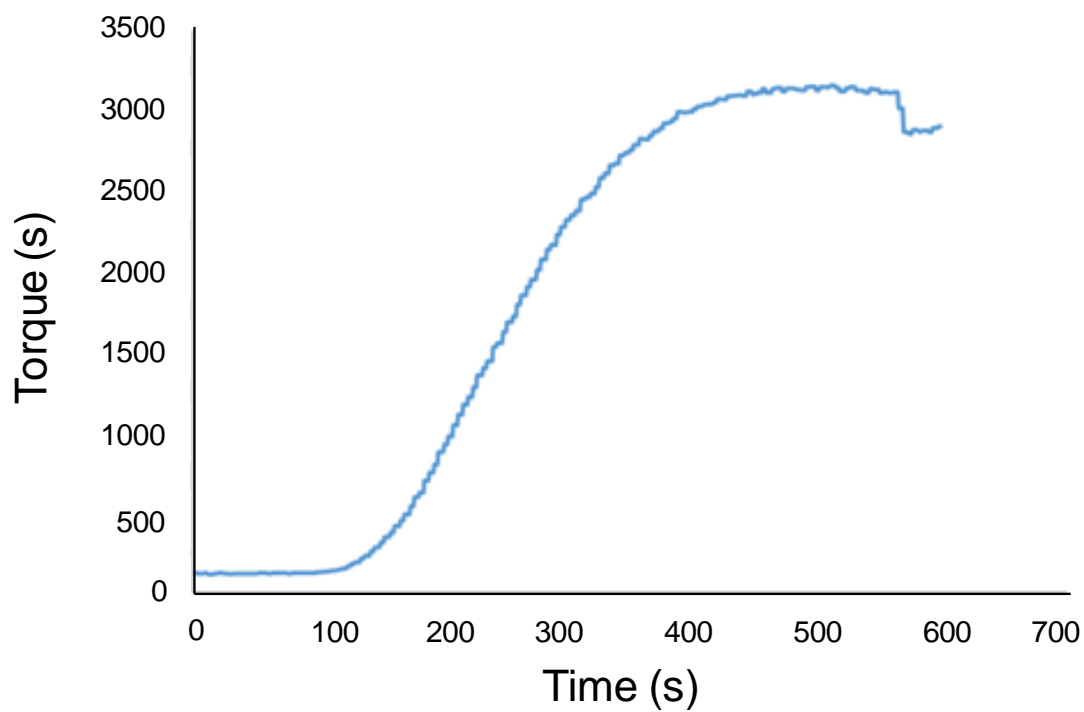

**Figure S3**. Torque as a function of time for the sample PLLA_DOD_R.

**Table S1**. Characteristics of the prepared samples

| Sample code | Reator type | $M_n$ [g·mol$^{-1}$] | $M_w$ [g·mol$^{-1}$] | MWD |
|---|---|---|---|---|
| PLLA_DOD_B | batch | 50600 | 77200 | 1.5 |
| PLLA_DOD_R | REX | 30800 | 44050 | 1.3 |

Table S2. Crystallinity of the prepared samples

| Sample code | Xm [%] | Xms [%] | Xm,total [%] |
|---|---|---|---|
| PLLA_DOD_R | 61 | - | 61 |
| PLLA_DOD_R_G | 62 | - | 62 |
| PLLA_Pyr_2500_R | 51 | 2 | 53 |
| PLLA_Pyr_2500_R_G | 49 | 7 | 56 |
| PLLA_Pyr_8000_R | 43 | 11 | 54 |
| PLLA_Pyr_8000_R_G | 50 | 8 | 58 |

The subscript $m_s$ indicates the value measured for the stereocomplexed fraction

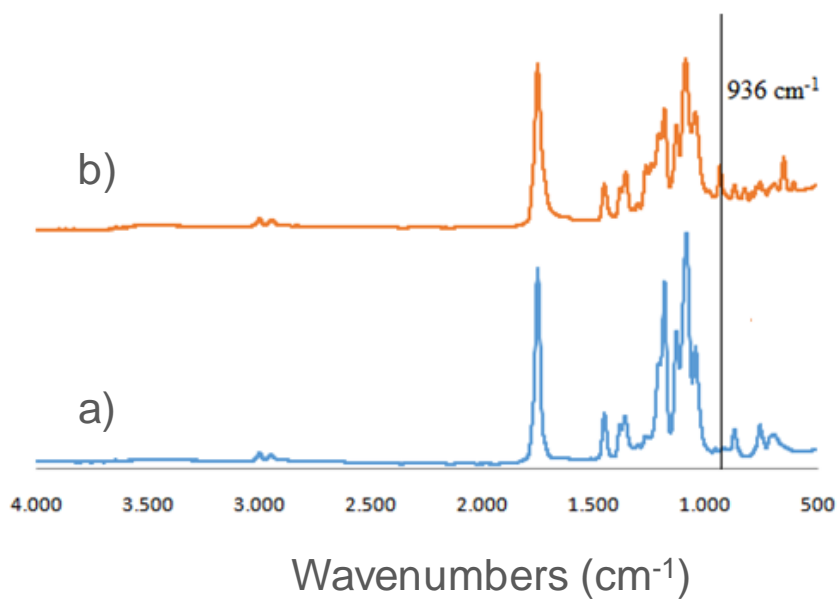

**Figure S4**. FTIR spectra of: a) PLLA_DOD_R and b) PLLA_DOD_B.

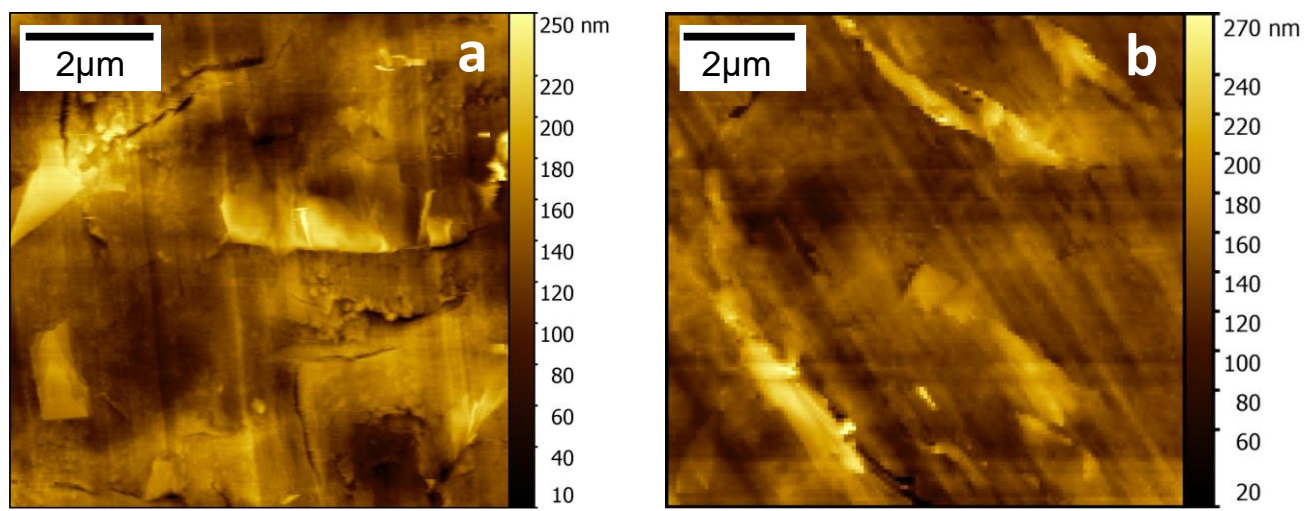

**Figure S5**. Representative topography images for dispersed GNP flakes within the polymer: a) PLLA_DOD_R_G and b) PLLA_Pyr_8000_R_G. in both cases, GNP flakes arising above the polymer surface, as a consequence of ultracryotomy, are clearly visible

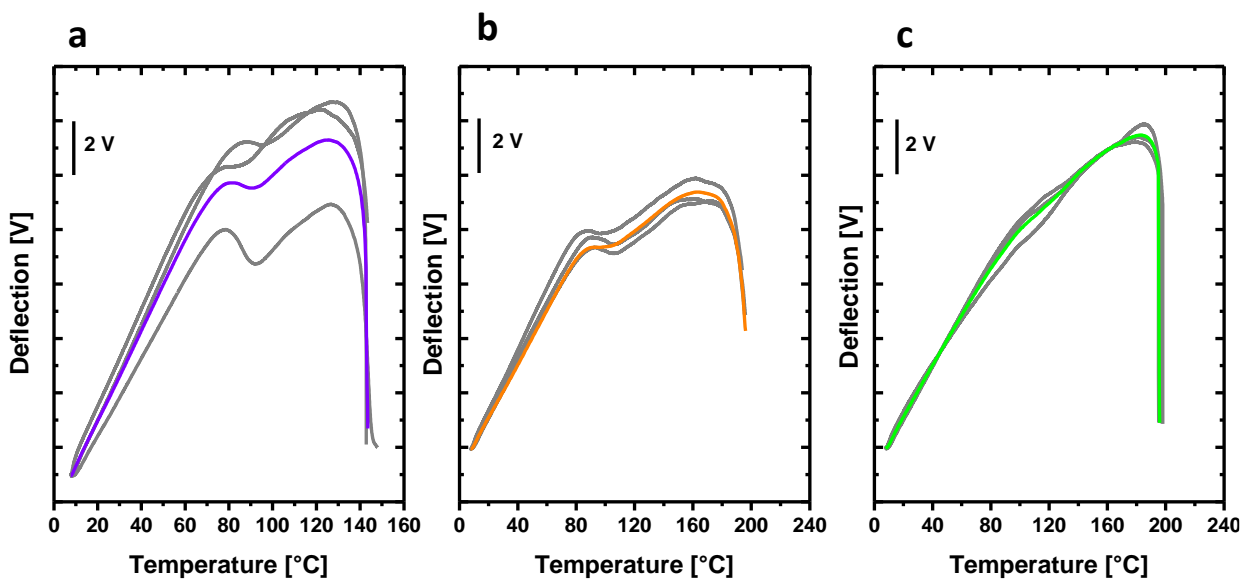

**Figure S6**. Nano-TA deflection vs temperature plots for PLLA (a), a mixture of PDLA and PLLA (1:2 ratio) forming 50% stereocomplex (b) and a mixture of PDLA and PLLA (1:1 ratio) forming 100% stereocomplex.

Typical NanoTA plots for PLA (Figure 5S) show increasing vertical deflection caused by thermal expansion of the sample until the temperature at which cold crystallization of PLA occurs above 80°C, leading to a reduction of vertical deflection of the cantilever. Once cold crystallization is completed, at temperature above 100°C, vertical deflection increases again as a consequence of further thermal expansion, until melting of the crystalline phase, corresponding to the penetration of the probe and related drop in the deflection curve, observed at 141±1°C. These two features in the deflection vs temperature curve result in a typical M-shaped plot, in agreement with a previous report. In the presence of PLA stereocomplexation, the shape of the plot is modified proportionally to the amount of stereocomplexation. In a mixture of PDLA and PLLA (1:2 ratio) forming 50% stereocomplex, the local minima in the deflection related to cold crystallization is significantly reduced while the temperature for the probe penetration is increased to 190±2°C. For fully stereo-complexed 1:1 PDLA and PLLA mixture, only traces of the low temperature feature assigned to cold crystallizazion are observed, while the average penetration temperature is observed at 192±4°C.